\documentclass[11pt]{article}
\usepackage{graphicx}
\usepackage{graphics}
\usepackage{epsfig}
\usepackage{amsmath}
\usepackage[dvips]{feynmp}
\newif\ifdimspec
\def\figsize#1{\dimspecfalse \checkdim#1\end
\ifdimspec
  \def\figureWidth{#1}%
\else
  \def\figureWidth{#1 in}\fi}
\def\checkdim#1{\ifx#1\end \let\next=\relax
  \else \ifcat#1a \dimspectrue \fi \let\next=\checkdim\fi \next}
\newcommand{\lblcaption  }[2]{\caption{#2\label{\secname#1}}}
\newcommand{\twoAngFiguresEPS}[6]
{
\begin{figure}
\begin{center}
\figsize{#4}
\begin{minipage}[t]{3.2in}
\begin{center}
\epsfig{file=\sectiondir/#1.eps,width=\figureWidth,angle=#5}
\end{center}
\end{minipage}
\begin{minipage}[t]{3.2in}
\begin{center}
\epsfig{file=\sectiondir/#2.eps,width=\figureWidth,angle=#6}
\end{minipage}
\end{center}
\lblcaption{#1}{#3}
\end{figure}
}
\newcommand{\BABARPubYear}    {01}

\newcommand{\BABARConfNumber} {03}

\input pubboard/babarsym
\def\BRbztodstdstNum {\ensuremath{(8.0 \pm 1.6(stat) \pm
1.2(syst))\times 10^{-4}}}
\def\BRbztodstdst {\ensuremath{{\BR}(\Bztodstdst) = \BRbztodstdstNum}}

\def\NSigData {38}
\def\NBkgData {\ensuremath{6.2 \pm 0.5}}
\def\fsideVal {1.72}
\def\fsideErr {0.10}

\def\TotLumi {23.3\invfb}
\def\OnResLumi {20.7\invfb}

\def\NBB {22.7 \times 10^{6}}

\def\DelELowSide {50}
\def\DelEHiSide {200\mev}
\def\mesLowSide {5.20}
\def\mesMidSide {5.26\gevcc}
\def\mesHiSide  {5.29\gevcc}

\def\Dstarp     {\ensuremath{D^{*+}}\xspace}
\def\Dstarm     {\ensuremath{D^{*-}}\xspace}
\def\Dp         {\ensuremath{D^+}\xspace}
\def\Dm         {\ensuremath{D^-}\xspace}

\def\BtoDDbar	{\ensuremath{B \to D^{(*)} \Dbar^{(*)}}\xspace}
\def\BtoDsDbar	{\ensuremath{B \to D^{(*)}_S \Dbar^{(*)}}\xspace}
\def\BztoDDbar	{\ensuremath{B^0 \to D^{(*)+} D^{(*)-}}\xspace}

\def\Dstptopip  {\ensuremath{\Dstarp \to \Dz\pip}\xspace}
\def\Dstptopiz  {\ensuremath{\Dstarp \to \Dp\piz}\xspace}

\def\DeltaE     {\ensuremath{\Delta E} \xspace}
\def\SsqovSpB   {\ensuremath{S^2/(S+B)}\xspace}

\def\chisqM     {\ensuremath{\chi^2_{Mass}}\xspace}

\long\def\inst#1{\par\nobreak\kern 4pt\nobreak
    {\it #1}\par\vskip 10pt plus 3pt minus 3pt}

\begin{document}
{\pagestyle{empty}

\begin{flushright}
\babar-CONF-\BABARPubYear/\BABARConfNumber \\
July, 2001 \\
\end{flushright}

\par\vskip 4cm

\begin{center}
\Large \bf Measurement of the branching fraction
for the decay \Bztodstdst 
\end{center}
\bigskip

\begin{center}
\large The \babar\ Collaboration\\
\mbox{ }\\
July 21, 2001
\end{center}
\bigskip \bigskip

\begin{center}
\large \bf Abstract
\end{center}
Decays of the type \BtoDDbar\ can be used to 
provide a measurement of the parameter
\stwob of the Unitarity Triangle that is complementary to the measurement
derived from the mode \bpsiks.  In this document we report a
measurement of the branching fraction 
for the decay \Bztodstdst\ with the
\babar\ detector.  With data corresponding to an integrated luminosity 
of \OnResLumi\ collected at the \FourS\ resonance
during 1999-2000, we have reconstructed \NSigData\ candidate signal events in
the mode \Bztodstdst\ with an estimated background of \NBkgData\
events.  From these events, we determine the branching fraction to be
\BRbztodstdst (preliminary).

\vfill
\begin{center}
Submitted to the\\ 20$^{th}$ International Symposium on Lepton and Photon
Interactions at High Energies, \\
7/23---7/28/2001, Rome, Italy
\end{center}

\vspace{1.0cm}
\begin{center}
{\em Stanford Linear Accelerator Center, Stanford University, 
Stanford, CA 94309} \\ \vspace{0.1cm}\hrule\vspace{0.1cm}
Work supported in part by Department of Energy contract DE-AC03-76SF00515.
\end{center}
}
\newpage

\begin{center}
\small

The \babar\ Collaboration,
\bigskip

B.~Aubert,
D.~Boutigny,
J.-M.~Gaillard,
A.~Hicheur,
Y.~Karyotakis,
J.~P.~Lees,
P.~Robbe,
V.~Tisserand
\inst{Laboratoire de Physique des Particules, F-74941 Annecy-le-Vieux, France }
A.~Palano
\inst{Universit\`a di Bari, Dipartimento di Fisica and INFN, I-70126 Bari, Italy }
G.~P.~Chen,
J.~C.~Chen,
N.~D.~Qi,
G.~Rong,
P.~Wang,
Y.~S.~Zhu
\inst{Institute of High Energy Physics, Beijing 100039, China }
% ...

\end{center}\newpage

\setcounter{footnote}{0}

\section{Introduction}
\label{sec:Introduction}
One of the most important goals of the \babar\ experiment is to
precisely measure the angles of the Unitarity Triangle.  While the
decay \bpsiks\ can be used to measure \stwob, the Standard
Model predicts that the time-dependent \CP violating asymmetries
in the decays~\cite{ref:cc}
\BztoDDbar can also
be used to measure the same quantity.  
An independent measurement of \stwob in
these modes would therefore provide a consistency
test of \CP-violation in the
Standard Model.

The vector-vector decay \Bztodstdst is not, however, a pure \CP\
eigenstate.  A sizeable dilution of the measured asymmetry may be
produced by a non-negligible $P$-wave \CP-odd component.  The
dilution can, in principle, be completely removed by a time-dependent angular
analysis of the decay products~\cite{ref:dunietz}.

The rate for the Cabibbo-suppressed decays \BtoDDbar can be estimated
from the measured rate of the Cabibbo-favored decays \BtoDsDbar:
\begin{equation}
{\BR}(\BtoDDbar) \approx
\left(\frac{f_{D^{(*)}}}{f_{D_S^{(*)}}}\right) \tan^2\theta_C
\cdot{\BR}(\BtoDsDbar),
\end{equation}
where $\theta_C$ is the Cabibbo angle, and $f_{D^{(*)}}$ and
$f_{D_S^{(*)}}$ are decay constants.
From this it follows that the \BtoDDbar branching fractions are of the
order of 0.1\%. Previous measurements of branching fractions and upper 
limits for these modes are summarized in Table \ref{prevbr}.

\begin{table}[bh]
\caption{\label{prevbr}Summary of branching fraction and upper limit
measurements performed by the CLEO experiment \cite{cleoprd62}. 
Upper limits are quoted at the 90\% confidence level.}
\begin{center} \begin{tabular}{lc}
\hline \hline
\multicolumn{1}{c}{Decay} & Branching Fraction ($\times 10^{-4}$) \\[0.5ex]
\hline
\Bztodstdst & $9.9^{+4.2}_{-3.3}(stat)\pm 1.2(syst)$ \\
\Bztodstd   & $<6.3$ \\
\Bztodd     & $<9.4$ \\
\hline \hline
\end{tabular} \end{center}
\end{table}

\section{The \babar\ detector and dataset}
\label{sec:babar}
The data used in this analysis were collected with the \babar\
detector~\cite{ref:babar} at the \pep2\ storage ring~\cite{ref:pepii}
located at the Stanford Linear Accelerator Center.  The results
presented in this paper are based on data taken during the 1999-2000
run.  This data sample represents an integrated
luminosity of \TotLumi, with \OnResLumi collected on the \FourS
resonance.  The total number of \BB pairs produced in this sample was
$N_{\BB} = \NBB$.

Charged particles are detected and their momenta measured with the
combination of a 40-layer drift chamber (DCH) and a five-layer silicon
vertex tracker (SVT) embedded in a 1.5\,T solenoidal magnetic field.
Photons are detected by a CsI electromagnetic calorimeter (EMC) that
provides excellent angular and energy resolutions with a high
efficiency for energies above 20\mev.  Charged particle
identification is provided by the specific ionization loss (\dedx)
in the tracking devices and by an internally reflecting ring-imaging
Cherenkov detector (DIRC) covering the barrel region of the detector.

\section{Determination of \BR(\Bztodstdst)}
\label{sec:Analysis}

\Bz mesons are exclusively reconstructed by combining two charged \Dstar
candidates reconstructed in a number of \Dstar and $D$ decay modes.
Events are pre-selected by requiring that there be
three or more charged tracks and that the normalized second Fox-Wolfram
moment~\cite{ref:fox} of the event be less than 0.6.  We also require
that the cosine of the angle between the reconstructed $B$ direction
and the thrust axis of the rest of the event be less than 0.9.

Charged kaon candidates are required to be inconsistent with the pion
hypothesis, as inferred from the Cherenkov ring
measured by the DIRC and the
\dedx as measured by the SVT and DCH.  
There are two exceptions to this: tighter kaon
identification is applied to one of the charged kaons in decay $D^+
\to \Km \Kp \pip$,
and no particle identification requirements are made
for the kaon from the decay
$\Dz \to \Km \pip$.

$\KS \to \pip\pim$ candidates are required to have an invariant mass
within 25\mevcc of the nominal \KS mass.  The opening angle between the
flight direction and the momentum vector of the \KS candidate is
required to be less than 200\mrad, and the transverse flight distance
from the primary event vertex must be greater than 2\,mm.

Neutral pion candidates are formed from
pairs of photons in the EMC with energy above 30\mev,
an invariant mass within 20\mevcc\ of the nominal \piz\ mass,
and a summed energy greater than 200\mev.
A mass-constraint fit is then applied to these \piz\ candidates. 
The \piz from \Dstptopiz decays (``soft'' \piz),
however, is required to have an invariant mass within
35\mevcc\ of the nominal \piz\ mass and momentum in the \FourS frame
of $70 < p^* < 450\mevc$, with no requirement on the summed
photon energy.

The decay modes of the \Dz and \Dp used in 
this analysis were selected
by an optimization of \SsqovSpB based on Monte Carlo simulations,
where $S$ is the expected number of signal events and $B$ is the
expected number of background events.  
The \Dz and \Dp modes used and their branching fractions are
summarized in Table~\ref{dzbr}.  \Dz (\Dp) meson candidates are required to
have an invariant mass within 20\mevcc of the nominal \Dz (\Dp) mass.

\begin{table}
\caption{\label{dzbr} \Dz and \Dp decay modes and branching
fractions~\cite{pdg}.  The branching fraction for $\KS \to \pipi$ is
included for modes containing a \KS.}
\begin{center} \begin{tabular}{lc}
\\ \hline \hline 
Decay Mode & Branching Fraction (\%) \\[0.5ex]
\hline
$\Dz \to \Km \pip$           & $3.83 \pm 0.09$ \\
$\Dz \to \Km \pip \piz$      & $13.9 \pm 0.9$  \\
$\Dz \to \Km \pip \pip \pim$ & $7.49 \pm 0.31$ \\
$\Dz \to \KS \pip \pim$      & $1.85 \pm 0.14$ \\
\hline
Total \Dz Branching Fraction & 27.1 \\
\hline \hline
Decay Mode & Branching Fraction (\%) \\[0.5ex]
\hline
$\Dp \to \Km \pip \pip$      & $9.0 \pm 0.6$   \\
$\Dp \to \KS \pip$           & $0.99 \pm 0.09$ \\
$\Dp \to \Km \Kp \pip$       & $0.87 \pm 0.07$ \\
\hline
Total \Dp Branching Fraction & 10.9 \\
\hline \hline
\end{tabular} \end{center}
\end{table}

The \Dstarp mesons
are reconstructed in their decays \Dstptopip and \Dstptopiz. 
We include for this analysis the decay combinations \Dstarp\Dstarm decaying 
to (\Dz\pip,\ \Dzb\pim) or (\Dz\pip,\ \Dm\piz), but not (\Dp\piz,\
\Dm\piz) due to 
the smaller branching fraction and larger expected backgrounds.
The branching fractions for these modes are summarized in Table \ref{dstarbr}.
\Dz and \Dp candidates are subjected to a mass-constraint fit and
then combined with soft pion candidates.  A vertex fit is performed
that includes the position of the beam spot to improve the angular
resolution of the soft pion.

\begin{table}
\caption{\label{dstarbr}\Dstar decay modes and branching fractions~\cite{pdg}.}
\begin{center} \begin{tabular}{lcc}
\hline \hline
Particle & Decay Mode & Branching Fraction (\%) \\[0.5ex]
\hline
\Dstarp & \Dstptopip & $67.7 \pm 0.5$ \\
        & \Dstptopiz & $30.7 \pm 0.5$ \\
Total Visible \Dstarp Branching Fraction & & 98.4 \\
\hline \hline
\end{tabular} \end{center}
\end{table}

To select \Bz candidates with well reconstructed \Dstar and $D$
mesons, we construct a $\chi^2$ that includes all measured \Dstar
and $D$ masses:

\begin{eqnarray*}
\chisqM =&
   \left(\displaystyle \frac{m_D - m_{D_{PDG}}}{\sigma_{m_D}}\right)^2
 + \left(\displaystyle \frac{m_{\Db} - m_{\Db_{PDG}}}{\sigma_{m_{\Db}}}\right)^2 \\
  &\quad + \left(\displaystyle\frac{\Delta m_{\Dstar} - \Delta
   m_{\Dstar_{PDG}}}{\sigma_{\Delta m}}\right)^2
 + \left(\displaystyle\frac{\Delta m_{\overline{D}^*} - \Delta
   m_{\Dstar_{PDG}}}{\sigma_{\Delta m}}\right)^2
\end{eqnarray*}
where the subscript $PDG$ refers to the nominal value, and $\Delta m$ 
is the $\Dstar - D$ mass difference.  For $\sigma_{m_D}$ we use
values computed for each $D$ candidate, while for $\sigma_{\Delta m}$
we use fixed values of 0.83\mevcc for \Dstptopip and 1.18\mevcc
for \Dstptopiz.  A requirement that $\chisqM < 20$ is applied to all \Bz
candidates.  In events with more than one \Bz candidate, we chose the
candidate with the lowest value of \chisqM.

A $B$ meson candidate is characterized by two kinematic variables.  We
use the energy-substituted mass, \mes, defined as
$$\mes \equiv \sqrt{{E_{Beam}^{*2}} - {p_B^*}^2}$$
and the difference of the $B$ candidate's energy from the beam energy,
\DeltaE,
$$\DeltaE \equiv E_{B}^* - E_{Beam}^* $$
where $E_{B}^*$ ($p_B^*$) are the energy (momentum) of the \B\ candidate
in the center-of-mass frame and $E_{Beam}^*$ is one-half of the
center-of-mass energy.
The signal region in the \DeltaE {\it vs.} \mes plane is defined to be
$|\DeltaE| < 25\mev$ and $5.273 < \mes < 5.285\gevcc$.
The width of this region corresponds to approximately $\pm 2.5\sigma$
in both \DeltaE and \mes.

To estimate the contribution from background in the signal region, we
define a sideband in the \DeltaE {\it vs.} \mes plane as
$$ |\DeltaE| < \DelEHiSide $$
$$ \mesLowSide < \mes < \mesMidSide $$
and
$$ \DelELowSide < |\DeltaE| < \DelEHiSide $$
$$ 5.26 < \mes < \mesHiSide $$
We parameterize the shape of the background in the \DeltaE {\it vs.}
\mes plane as the product of an ARGUS
function~\cite{ref:argus} in \mes and a first-order polynomial in \DeltaE.
Based on this parameterization we estimate that the ratio
of the number of background events in the signal region
to the number in the sideband region is
$(\fsideVal \pm \fsideErr)\times 10^{-2}$.
The uncertainty is derived from the observed variation of this ratio
under alternative assumptions 
for the background shape in \mes\ and \DeltaE.

\begin{figure}[!htb]
\begin{center}
\includegraphics[height=7cm]{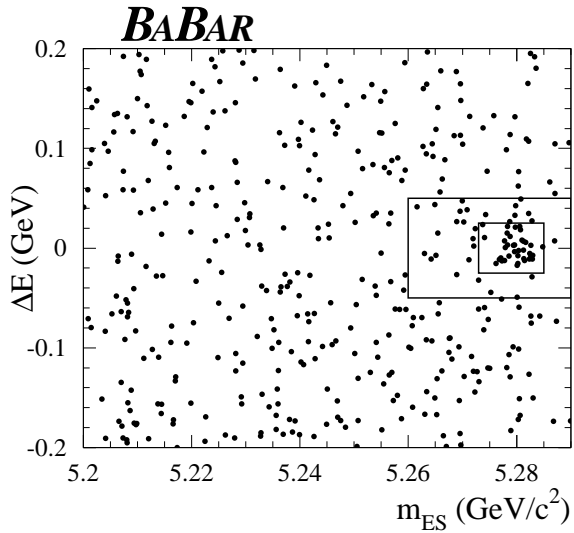}
\caption{Distribution of events in the \DeltaE {\it vs.} \mes plane.
The small box indicates the signal region, while the sideband region
is everything outside the larger box.
}
\label{fig:mesdeltae}
\end{center}
\end{figure}

\begin{figure}[!htb]
\begin{center}
\includegraphics[height=7cm]{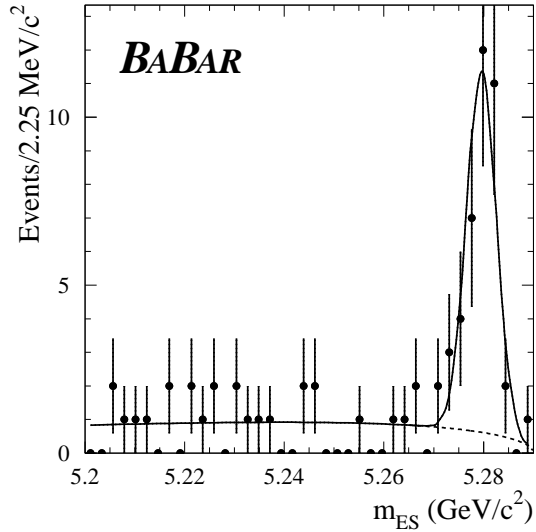}
\caption{Distribution of events in \mes plane with a cut of $|\DeltaE|
< 25\,\mev$ applied.  The curve represents a fit to the distribution
of the sum of a Gaussian to model the signal and an ARGUS 
function~\cite{ref:argus} to model the background shape.
}
\label{fig:mes}
\end{center}
\end{figure}

Figure~\ref{fig:mesdeltae} shows the events in the \DeltaE {\it vs.}
\mes plane after all selection criteria have been applied.  The small
box in the figure indicates the signal region defined above, and the
sideband is the entire plane excluding the region bounded by the
larger box outside the signal region.  There are a total of 38 events
located in the signal region, with 363 events in the sideband region.
The latter, together with the effective ratio of areas of the signal
region to the sideband region, implies an expected number of
background events in the signal region of $6.24 \pm 0.33(stat) \pm
0.36 (syst)$.
The quoted systematic
uncertainty comes from the
background shape variation discussed previously.  Figure~\ref{fig:mes}
shows a projection of the data on to the \mes axis after requiring
$|\DeltaE| < 25\mev$.
 
We use a detailed Monte Carlo simulation of the \babar\ detector to
determine the efficiency for reconstructing the signal.  This,
together with the total number of \BB pairs produced during data
collection, allows us to determine a preliminary branching fraction for
\Bztodstdst to be

$$\BRbztodstdst$$

The dominant systematic uncertainty in this measurement comes from our
level of understanding of the charged particle tracking efficiency (9.4\%).
The high charged particle multiplicity in this decay mode makes this
measurement particularly sensitive to tracking efficiency. Uncertainties 
were assigned on a per track basis for $\pi$, $K$ and slow $\pi$,
 and were added linearly due to large correlations.  The imprecisely
known partial-wave content of the \Bztodstdst final state
is another source of systematic uncertainty (6.6\%).  This was estimated by
calculating the change in the reconstruction efficiency for different 
final angular states in Monte Carlo.
Other significant systematic uncertainties arise due to the
uncertainties on the ${\Dstar}^+$, \Dz and
$D^+$ branching fractions (5.6\%) and the differences in mass
resolutions between Monte Carlo and data (4.1\%). 
 The total systematic uncertainty from all sources is 14.5\%.


%
%


\section{Summary}
\label{sec:Summary}
Using data collected by the \babar\ experiment during 1999-2000, we
have observed a signal of $31.8 \pm 6.2(stat) \pm 0.4(syst)$ events in
the decay \Bztodstdst.  We measure a preliminary branching ratio to be
$$\BRbztodstdst$$

\section{Acknowledgments}
\label{sec:Acknowledgments}


We are grateful for the 
extraordinary contributions of our \pep2\ colleagues in
achieving the excellent luminosity and machine conditions
that have made this work possible.
The collaborating institutions wish to thank 
SLAC for its support and the kind hospitality extended to them. 
This work is supported by the
US Department of Energy
and National Science Foundation, the
Natural Sciences and Engineering Research Council (Canada),
Institute of High Energy Physics (China), the
Commissariat \`a l'Energie Atomique and
Institut National de Physique Nucl\'eaire et de Physique des Particules
(France), the
Bundesministerium f\"ur Bildung und Forschung
(Germany), the
Istituto Nazionale di Fisica Nucleare (Italy),
the Research Council of Norway, the
Ministry of Science and Technology of the Russian Federation, and the
Particle Physics and Astronomy Research Council (United Kingdom). 
Individuals have received support from the Swiss 
National Science Foundation, the A. P. Sloan Foundation, 
the Research Corporation,
and the Alexander von Humboldt Foundation.


\begin{thebibliography}{99}
\bibitem{ref:cc}
Charge-conjugate states are implied throughout this paper and the symbol
$D^{(*)}$ refers to either $D$ or $D^*$.
\end{thebibliography}
\end{document}